\documentclass[12pt,preprint]{aastex}
\usepackage{amsmath}

\begin{document}
\title{ON THE ORIGINS OF THE HIGH-LATITUDE H$\alpha$ BACKGROUND}
\author{Adolf N. Witt\altaffilmark{1}, Benjamin Gold\altaffilmark{2},  Frank. S. Barnes III\altaffilmark{3}, Casey T. DeRoo\altaffilmark{4}, Uma P. Vijh\altaffilmark{1}, \& Gregory J. Madsen\altaffilmark{5}}
\altaffiltext{1}{Ritter Astrophysical Research Center, University of Toledo, Toledo, OH 43606}
\altaffiltext{2}{Department of Physics \& Astronomy, The Johns Hopkins University, 3400 N. Charles St. Baltimore, MD 21218}
\altaffiltext{3}{Twin Oaks Observatory, Rock Hill, SC 29730}
\altaffiltext{4}{Concordia College, 901 8th St. S, Moorhead, MN 56562}
\altaffiltext{5}{Sydney Institute for Astronomy, School of Physics, University of Sydney, NSW 2006}

\begin{abstract}
The diffuse high-latitude H$\alpha$ background is widely believed to be predominantly the result of in-situ recombination of ionized hydrogen in the warm interstellar medium of the Galaxy. Instead, we show that both a substantial fraction of the diffuse high-latitude H$\alpha$ intensity in regions dominated by Galactic cirrus dust and much of the variance in the high-latitude H$\alpha$ background are the result of scattering by interstellar dust of H$\alpha$ photons originating elsewhere in the Galaxy. We provide an empirical relation, which relates the expected scattered H$\alpha$ intensity to the IRAS 100 $\mu$m diffuse background intensity, applicable to about 81\% of the entire sky. The assumption commonly made in reductions of CMB observations, namely that the observed all-sky map of diffuse H$\alpha$ light is a suitable template for Galactic free-free foreground emission is found to be in need of reexamination.
\end{abstract}

\section{INTRODUCTION}
Efforts to observe the all-sky distribution of diffuse H$\alpha$ emission (Dennison et al. 1998, VTSS \footnote{\url{http://www.phys.vt.edu/$\sim$halpha}}; Gaustad et al. 2001, SHASSA\footnote{\url{http://amundsen.swarthmore.edu/SHASSA}}; Reynolds et al. 2002, Haffner et al. 2003, WHAM\footnote{\url{http://www.astro.wisc.edu/wham}}) have resulted in detailed maps, which revealed the presence of diffuse, highly structured H$\alpha$ radiation at high Galactic latitudes. Although questions about the inferred ionization sources for diffuse, ionized gas outside of H II regions in late-type galaxies have not been fully resolved \citep[e.g.][]{Ferguson+96, HW00, Seon09}, the high-latitude H$\alpha$ emission in the Milky Way has generally been interpreted as resulting from in-situ emission in an extended, ionized warm interstellar medium (WIM) \citep{Gaensler+08, Haffner+09}. This was supported by studies of the [S II]/H$\alpha$ ratio in a limited number of selected regions \citep{Reynolds85, Reynolds88, Haffner+03, MRH06} and with radiative transfer models by \citet{WR99}, who concluded that scattering of H$\alpha$ photons originating elsewhere in the Galaxy by high-latitude dust was not a significant component of the H$\alpha$ background. Consequently, an all-sky H$\alpha$ map by \citet{Finkbeiner03}, produced through the combination of the VTSS, SHASSA, and WHAM data, has been used extensively as a template for Galactic free-free emission in the analysis of Galactic foregrounds of the all-sky CMB observations by WMAP \citep{RH05, Bennett+03, DDD03, Gold+10}, with the fundamental assumption that the extinction-corrected H$\alpha$ intensity is directly proportional to the line-of-sight emission measure in the foreground Galactic interstellar medium.

The question of which physical processes are responsible for high-latitude H$\alpha$ radiation was reopened recently through the analysis of the H$\alpha$ excess observed in the isolated, high-latitude interstellar cloud LDN 1780 (l = $359\arcdeg.02$; b = $+36\arcdeg.77$; distance $\sim 110$ pc; \citet{Franco89}). LDN 1780 is readily identifiable as a distinct object on the map of \citet{Finkbeiner03} by its excess H$\alpha$ surface brightness. From an analysis of the H$\alpha$ morphology, in addition to CO, FIR, and optical extinction maps of LDN 1780, \citet{d-BC06} concluded that the observed surface brightness distribution of LDN 1780 was inconsistent with photo-ionization as the cause for the observed H$\alpha$ light. Photoionization by an external field of Lyman continuum photons is expected to result in ionization of a thin outer shell of an isolated interstellar cloud while the observations of the H$\alpha$ light from LDN 1780 clearly reveal the dense inner core of the cloud as the brightest feature (see Fig. 1(bottom)), showing that most of the observed H$\alpha$ photons come from the densest interior structures of the cloud. To account for this phenomenon, \citet{d-BC06} proposed that cosmic rays caused partial ionization of the hydrogen throughout LDN 1780, leading to the observed H$\alpha$ emission. To produce the observed intensity, they postulated a locally enhanced cosmic ray flux about one order of magnitude larger than commonly accepted for the solar neighborhood. 

\citet{MJL07}, analyzing the same LDN 1780 data from the \citet{Finkbeiner03} map as \citet{d-BC06}, concluded instead that the H$\alpha$ excess intensity in LDN 1780 and a number of other similar high-latitude clouds is more likely the result of scattering by the dust in these clouds of H$\alpha$ photons produced elsewhere in the Galaxy. Their interpretation relies upon the facts that (1) the optical depth of LDN 1780 at the wavelength of H$\alpha$ is sufficiently small so that external H$\alpha$ photons from the Galactic interstellar radiation field (ISRF) can readily reach the inner structures of LDN 1780 and (2) the known albedo of interstellar dust is sufficiently high to produce the observed intensities by scattering, given the known mean intensity of the Galactic radiation field in the H$\alpha$ line. Still more recently, \citet{LJM10} analyzed the excess H$\alpha$ surface brightness of the large high-latitude translucent cloud G294-24 and concluded that also there the excess H$\alpha$ intensity can be explained solely by galactic H$\alpha$ radiation scattered off dust in the cloud, thus supporting the conclusions of \citet{MJL07}.

The possible presence of scattered H$\alpha$ radiation at high latitudes was first considered by \citet{Jura79}, but subsequent Monte Carlo radiative transfer simulations by \citet{WR99} suggested that the scattered H$\alpha$ component should be relatively small. They conclude that the scattered light from midplane H II regions is typically around 10\% of the total H$\alpha$ intensity at high latitudes, with a similar fraction (see their Fig. 3) being contributed by scattering of H$\alpha$ photons emitted by the WIM. Thus, their models predict an average scattered H$\alpha$ contribution at high latitudes of about 20\% of the total H$\alpha$ intensity. Obvious shortfalls of the Wood \& Reynolds model are their assumption of a smooth, unstructured ISM, and more seriously, the use of the \citet{HG41} scattering phase function. The Henyey-Greenstein phase function is known to under-predict scattered light intensities \citep[e.g.][]{Witt77, Draine03}, when large scattering angles are involved as they are in the scattering of midplane H$\alpha$ photons by high-latitude clouds.

Frequently cited support for the conclusion that scattering makes only a minor contribution to the diffuse H$\alpha$ background comes from WHAM measurements of the [S II]/H$\alpha$ ratio cited earlier in this section, which demonstrated that this ratio is typically several times larger in the diffuse ionized interstellar medium than in bright H II regions. If scattering of H$\alpha$ photons originating in H II regions were to dominate the H$\alpha$ intensity in the diffuse ISM, the [S II]/H$\alpha$ ratio would be expected to be similar to that in H II regions. We will show in this paper that the selection criteria in use for the [S II]/H$\alpha$ ratio measurements conducted with WHAM have led to a highly biased sample of regions in which scattering indeed is a minor contributor. The [S II]/H$\alpha$ argument, while valid in principle, has been inappropriately extended to the entire sky without sufficient observational justification.

Given the ubiquity of high-latitude dust structures in the Galactic cirrus \citep{Sandage76, Low+84}, the conclusions of \citet{MJL07} and \citet{LJM10} have profound consequences. If a substantial fraction of the high-latitude H$\alpha$ background is due to scattering rather than in-situ recombination, be that in balance to photo-ionization \citep{Reynolds+95} or cosmic ray ionization as proposed by \citet{d-BC06}, the template used for Galactic free-free emission in the WMAP analyses \citep{Bennett+03, Gold+10} employed systematically intensities that were too high, as well as spatial brightness variations not actually present in the distribution of the in-situ emission component alone. 

In this paper we present new observations of LDN 1780 in H$\alpha$ and in adjacent continuum bands with superior spatial resolution and sensitivity compared to the \citet{Finkbeiner03} data. We use these new data to independently examine the case for dust scattering in LDN 1780. Scattering by dust would clearly be the phenomenon with greater consequences, because it does not require special conditions such as the locally, ten-fold enhanced cosmic ray flux postulated by \citet{d-BC06}. With the case for scattering made by \citet{MJL07} and by \citet{LJM10} confirmed, we then extend our conclusions to the entire high-latitude sky with consequences for future CMB analyses.
We describe our new observations in Sect. 2, and discuss the calibrations and reductions in Sect. 3. In Sect. 4 we analyze the excess H$\alpha$ intensity found in LDN 1780. Sect. 5 contains the discussion of the consequences of our results for the high-latitude H$\alpha$ background in general. Our conclusions are summarized in Sect. 6.

\section{OBSERVATIONS}

We used two identical, remotely operated Takahashi Epsilon 180ED astrographs (f/2.8) to obtain new imaging data for LDN 1780. The telescopes were located at New Mexico Skies Observatory near Mayhill, NM, at an altitude of 7300 ft (2225 m). Each telescope was equipped with a Santa Barbara Imaging Group STL-6303E CCD camera, covering a field of $2\arcdeg \times 3\arcdeg$. The angular resolution determined by the 9 $\mu$m pixels was 3.70$\arcsec$ px$^{-1}$. Each camera was equipped with an 8-position filter wheel, jointly capable of housing 15 BATC intermediate-band filters \citep{Yan+00} plus a narrow-band Astrodon H$\alpha$ filter. The filter characteristics, total exposure times, and observing dates are summarized in Table 1. The data were part of a larger program directed at studying the optical spectral energy distributions of high-latitude clouds in conjunction with Spitzer infrared band observations with both the MIPS and IRAC imaging systems (Witt et al. 2010, in preparation).

The observations were carried out under photometric conditions. Both astrograph/camera systems were flux-calibrated through observations of the absolute flux standard star SAO 90153. For LDN 1780, individual sub-exposures of 900 s each were processed through bias subtraction, flat-fielding and cosmic ray removal. Atmospheric extinction corrections were applied to each 900 s sub-exposure, before co-adding them. This was particularly important in the case of LDN 1780, because its negative declination required data acquisition over a fairly large range of air masses. We note here that the SHASSA data \citep{Gaustad+01}, which form the basis for the \citet{Finkbeiner03} H$\alpha$ map for LDN 1780 were not corrected for atmospheric extinction.

The narrow-band H$\alpha$ filter has a nearly rectangular transmission profile with 6 nm FWHM. In addition to the H$\alpha$ line, this filter also transmits scattered-light continuum around H$\alpha$ and possibly light from the [N II] line at 654.81 nm and to a lesser extent light from the [N~II] line at 658.34 line, if there is significant radiation present in these lines in LDN 1780. As discussed by \citet{MRH06}, these lines are fairly weak relative to H$\alpha$ in H~II regions but are more significant in the radiation from the WIM. Observations of stellar continuum sources with spectral energy distributions similar to that of LDN 1780 with the BATC \#9 filter and the H$\alpha$ filter were used to determine the continuum contribution passing through the finite width (6 nm) of the H$\alpha$ filter. 

On 13 July and 14 July, 2010, one of us (GJM) used the Wisconsin H-Alpha Mapper (WHAM), currently located at CTIO, to measure the emission line ratios [N II]/H$\alpha$ for the stronger [N~II] line at 658.34 nm and [S II]/H$\alpha$ for the [S~II] line at 673.1 nm on LDN 1780 relative to the nearby sky. The 1-degree WHAM beam\footnote{http://www.astro.wisc.edu/wham/old/description.html} encloses the full extent of LDN 1780 as well as as substantial fraction of the adjacent sky. The excess H$\alpha$ intensity of LDN 1780, averaged over the 1-degree beam relative to three sky positions at several degrees offset (same latitude, $1\arcdeg.52$ to lower l, $1\arcdeg.30$ to higher l; same longitude, $3\arcdeg.0$ to higher b) was found to be 0.90 R; the H$\alpha$ line profile was substantially narrower (18 km/s) than commonly seen in the high-latitude WIM (25 - 30 km/s). The line ratios were measured as [N II]/H$\alpha$ = 0.19 and [S II]/H$\alpha$ = 0.16, rather close to corresponding values in H II regions and much smaller than typically found in the diffuse WIM \citep{MRH06}.

While these recent WHAM measurements suggest that significant contamination of our H$\alpha$ intensities on LDN 1780 by contributions from the [N~II] lines is not likely, we will carry out an additional examination of this question in Sect. 3 by comparing our measured intensities with those derived from the SHASSA data, which have been calibrated with pure WHAM H$\alpha$ intensities and should therefore be free of [N II] contamination. In the absence of contamination, we expect the two sets of H$\alpha$ intensities to agree within the measurement uncertainties.

\section{ADDITIONAL DATA AND REDUCTIONS}
\subsection{Surface Brightness Morphology}
We conducted differential surface brightness measurements for 378 individual circular source regions on LDN 1780 (radius $37\arcsec$) and the adjacent sky, represented by the average of 40 circular sky fields, as displayed in Fig. 1(top). The positions of the source regions were selected by eye to avoid contributions from stars as faint as V $\sim 21$\ mag, yet were aimed to provide maximum surface coverage of the visible extent of LDN 1780. 

The source regions were divided into five color-coded groups to distinguish regions by the dust column density through the cloud and the ISRF impinging upon the dust in them (Fig. 1(top)). Fig. 1(center) displays the image of LDN 1780 in the narrow-band H$\alpha$ filter. The morphology of LDN 1780 observed with the H$\alpha$ narrow-band filter appears to be identical to that in adjacent continuum filters. Given the width (6 nm) of the H$\alpha$ narrow-band filter, both H$\alpha$ line photons as well as continuum photons are contributing to this image. Fig. 1(bottom) displays the residual H$\alpha$ image of LDN 1780, after the data in Fig. 1(center) have been sky-subtracted and corrected for the continuum contribution in the narrow-band H$\alpha$ filter. The latter was determined to be 12.38\% of the continuum intensity measured through the broader BATC \#9 filter, normalized to the same total exposure time, sky-background and extinction corrected. This image shows the excess H$\alpha$ of LDN 1780 only. In order to overcome the large per pixel noise but still leave the stars as recognizable discrete sources, the image has been smoothed by a 7-pixel Gaussian. The most important aspect of the resulting image is that the high-opacity inner core region (A$_V \sim 4$ mag; \citet{Ridderstad+06}) of LDN 1780 appears as a region of greatest H$\alpha$ brightness. The implications of this fact will be discussed below in Sect. 4.

\subsection{External Data}
Other data sets employed in this study were the all-sky H$\alpha$ map of \citet{Finkbeiner03} and the all-sky maps of the IRAS 100 $\mu$m diffuse background and the line-of-sight dust extinction maps of \citet{SFD98}. Finally, we used numerical data for the optical depths through LDN 1780 that were derived from the intensity at 160 $\mu$m (Spitzer, MIPS) and the dust temperatures determined by Witt et al. (2010, in preparation). The dust temperatures were determined from the intensity ratios at 160 $\mu$m (MIPS) and 100 $\mu$m (IRAS) and thus refer to the equilibrium temperature of the large-grain population responsible for the far-infrared emission as well as the optical scattering in LDN 1780. The temperatures of these grains were found to lie in the range 14.5 K to 16.8 K, typical for dust in interstellar clouds exposed to the ISRF near the sun \citep{Bot+09} and in good agreement with the dust temperatures for LDN 1780 independently determined by \citet{Ridderstad+06}. We note that these dust temperatures are well below the minimum temperature (17.35 K) indicated by the \citet{SFD98} dust temperature map for the center of LDN 1780. The reason for this discrepancy most likely arises from the low spatial resolution of the COBE/DIRBE 100 $\mu$m and 240 $\mu$m maps, which are the basis for the Schlegel et al. temperature map. With our temperatures based on the much higher IRAS 100 $\mu$m resolution, we determined the optical depth at 160 $\mu$m for each sampling position on LDN 1780 from the relation $\tau_{160} = I_{160}/ B_{160}(T)$, where $B_{160}(T)$ is the Planck function for the dust temperature at the specific sampling field and $I_{160}$ is the corresponding measured MIPS intensity, both at the wavelength of 160 $\mu$m.

\subsection{Possibility of Emission Line Contamination}
We examined our measurements of the continuum-corrected excess H$\alpha$ intensities with regard to the reliability of our absolute calibration and the possible presence of [N II] line contamination by comparing them with the corresponding measurements of the excess intensities in the Finkbeiner H$\alpha$ map. The two data sets differ in several important aspects. (1) Resolution: Our data (NMS, for New Mexico Skies) have a scale of 3.70$\arcsec$ px$^{-1}$ while the SHASSA survey, which underlies this region of the Finkbeiner H$\alpha$ map, was conducted with an angular scale of 47.64$\arcsec$ px$^{-1}$, with the final resolution of the \citet{Finkbeiner03} map reduced to a resolution of $6\arcmin$; (2) Star subtraction: Our data are free of starlight contamination to the limit of V $\sim 21$\ mag by design while the SHASSA survey relied upon star subtraction based on stellar images in adjacent continuum images, which was not perfect in all cases (see \citet{Gaustad+01} and \citet{Finkbeiner03} for details); (3) Atmospheric extinction correction: The NMS data were corrected for atmospheric extinction at the sub-exposure (900 s) level while the SHASSA data where not corrected for atmospheric extinction; and (4) Absolute calibration: The NMS data were calibrated against the absolute flux standard star SAO 90153, whereas the SHASSA data were calibrated against sources with known H$\alpha$ fluxes, with an absolute zero point based on WHAM H$\alpha$ measurements, which are pure, uncontaminated H$\alpha$ intensities. The LDN 1780 data obtained from the Finkbeiner map should, therefore, be free of contamination from [N II] lines, while the NMS data may not be.

The comparison of the intensities measured in LDN 1780 is shown in Fig. 2. In order to facilitate this comparison, the resolution of the Finkbeiner map was artificially increased to match that of the NMS data. The ideal 1:1 correlation is indicated by the solid diagonal line; the least-squares linear fit (dashed line) through the origin suggests that the Finkbeiner intensities, on average, are about 11\% higher than the corresponding NMS intensities. This difference is small compared to the typical errors in each measurement represented by the error bars in Fig. 2. This independent test is consistent with our recent measurement of the [N II]/H$\alpha$ line intensity ratio for LDN 1780 (see Sect. 2), which showed the [N II] lines to be weak. We, therefore,  conclude that contamination of the NMS data by light from [N II] lines passing through our H$\alpha$ filter is insignificant.

\section{ANALYSIS OF THE H$\alpha$ EXCESS IN LDN 1780}
As seen in Fig. 1(bottom), LDN 1780 stands out brightly against the surrounding sky at H$\alpha$. We further observe that both sets of measurements shown in Fig. 2 consistently identify the densest parts of LDN 1780 with A$_V \sim 4$ mag, represented by the red symbols (see Fig. 1(top)), as the region of maximum H$\alpha$ intensities. Such brightness contrast would not be expected, if the H$\alpha$ radiation from LDN 1780 were the result of photo-ionization. If ionizing photons could penetrate via low-density paths and voids through a turbulent interstellar medium to the altitude of LDN 1780, as suggested recently by \citet{Wood+10}, the resulting ionization would be limited to a thin outer shell of LDN 1780, given its maximum visual extinction of A$_V \sim 4$ mag and its known nature as a molecular cloud with a dense core and a more diffuse outer envelope \citep{Toth+95}. In such a morphology, the outer, optically thin regions would appear brightest in H$\alpha$ line emission, while the optically thick core regions would appear only about half as bright, as the extinction in the core would attenuate the H$\alpha$ radiation from the far side. This is contrary to the observations shown in Fig. 1(bottom). The H$\alpha$ morphology of LDN 1780 is indistinguishable from its appearance in the adjacent continuum, which is well understood as originating through scattering of the ISRF by the dust grains in LDN 1780 \citep{Mattila79}. These facts strongly imply that photoionization by external Lyman continuum photons is not the cause of the observed H$\alpha$ light in LDN 1780, while dust scattering of H$\alpha$ photon from elsewhere in the Galaxy could provide a natural explanation.

The same reasoning caused \citet{d-BC06} to look for an ionization source other than photoionization for LDN 1780, one that would be capable to penetrate deeply into the interior of the cloud. They suggested ionization by cosmic rays, which would be supported by a linear relation between the resulting H$\alpha$ intensity and the optical depth through the cloud. In our view, they were misled into believing that there exists such a linear relation between the H$\alpha$ intensity and the visual extinction through LDN 1780 by applying what we consider an inappropriate extinction correction (their equation (7)). We will show in this section that the actual observed functional dependence of the H$\alpha$ surface brightness on the optical depths of different cloud regions is a natural consequence of the radiative transfer of an external ISRF through the interior of LDN 1780.

The actual observed scattering response of LDN 1780 to illumination by an external ISRF in terms of the intensity $I$ as a function of the line-of-sight optical depth $\tau$ is the solution to the equation of transfer. In the range of optical depths from the optically thin limit to moderately optically thick regimes as represented by LDN 1780, this solution has the general functional form:

\begin{equation}
I(\tau)  =  A [1-exp (-\tau)]
\end{equation}			

The constant $A$ is a product of three quantities: (1) the average intensity $J$ of the incident ISRF; (2) the dust albedo $a$; and (3) the phase function contribution factor $P$, which measures the coupling between the anisotropic external ISRF to the line of sight to the cloud through the forward-directed scattering phase function of the dust grains \citep{Witt85}. Rather than trying to find the solution to the actual transfer problem for LDN 1780 through modeling with all its inherent approximations and uncertainties, we will employ an empirical approach by using the well-observed scattered light intensities in our two continuum filters  immediately adjacent to H$\alpha$, BATC \#8 (607 nm) and BATC \#10 (705 nm), to provide the functional parameters for the solution to the equation of transfer.

In Fig. 3a we plot the averaged observed intensities $I_{cont}$ in these two filters against the optical depth $\tau_{160}\ \mu$m (see Sect. 3) for all observed positions on LDN 1780. The least-squares curve fit shown corresponds to the function:

\begin{equation}
I_{cont}  =  A [1 - exp( -B \tau_{160})],		
\end{equation}

\noindent where the constant $A$ is the appropriate value for the continuum adjacent to H$\alpha$ and the constant $B$ is the ratio of the optical depths around H$\alpha$ and at 160 $\mu$m. We find these values: $A = 31.86 \times 10^{-9}\ \mathrm{erg\ cm^{-2}\ s^{-1}\ \AA^{-1}\ sr^{-1}}$ and $B = 1157.92$. This lets us conclude that the maximum optical depth of LDN 1780 near H$\alpha$ is $\tau_{H\alpha} = 0.0026 \times 1157.92 = 3.01$, consistent with the maximum extinction through LDN 1780 of $A_V$ = 4.2 mag found by \citet{Ridderstad+06} and the ratio of $\tau_{H\alpha} / A_V = 0.714$ \citep{WD01} for Milky Way dust with $R_V = 3.1$. The average continuum ISRF intensity in the vicinity of the sun near H$\alpha$ is estimated to be $J(656\ nm) = 1.5  \times 10^{-7}\ \mathrm{erg\ cm^{-2}\ s^{-1}\ \AA^{-1}\ sr^{-1}}$ \citep{PS05} and the dust albedo is modeled as $a = 0.56$ \citep{Draine03}. This allows us to estimate the phase function contribution factor $P = 0.38$. A value of $P$ less than unity is expected for forward-scattering grains and a scattering geometry where large-angle scattering prevails, as is the case of high-latitude clouds illuminated preferentially by sources near the Galactic plane \citep{Witt85, Stark95}. As Fig. 3a shows, the ISRF impinging upon LDN 1780 is clearly non-isotropic. The southern regions facing the Milky Way plane, represented by blue, cyan and green symbols are illuminated by a stronger radiation field, whereas the northern regions represented by the yellow symbols are illuminated by intensities that are 20\% to 30\% lower compared to the southern regions. 

A relation identical in functional form to Eq. (2) should also fit the measurements of the H$\alpha$ excess surface brightness, if the H$\alpha$ light in LDN 1780 is a result of scattering by the same grains with the same scattering geometry as the ones that apply to the adjacent continuum. To arrive at the appropriate value for the constant $A$, the first factor representing the average continuum radiation density needs to be replaced by the average H$\alpha$ intensity. This is found by averaging the Finkbeiner H$\alpha$ map over the entire sky. \citet{MJL07} and \citet{LJM10} report a value for the all-sky H$\alpha$ average intensity of $8\ R$. The other two factors contributing to the value of the constant A remain  approximately the same as found earlier for the continuum, given that the illumination geometries for the continuum and H$\alpha$ are very similar. Thus, we estimate $A(H\alpha) = 1.70\ R$, if we assume that the ratio of the H$\alpha$ line intensity to continuum intensity as seen by LDN 1780 is the same as found near the sun. The constant $B$ should be identical to that found from Fig. 3a, because, by design, $B$ refers to an effective wavelength of 656 nm, the same as H$\alpha$.

We plot our H$\alpha$ surface brightness measurements as a function of the optical depth $\tau_{160}$ in Fig. 3b. The solid curve is the expected scattering solution for $A = 1.70\ R$. It fits the H$\alpha$ intensities for the northern portion of LDN 1780 rather well but fails to predict the higher intensities in the southern and central portions of the cloud. A more satisfactory solution for these portions of LDN 1780 with $A = 2.35$ is represented by the dashed curve, which implies that the radiation field as seen by the southern side of LDN 1780 has a 38$\%$ higher ratio of H$\alpha$ to continuum intensities than the ISRF near the sun. An examination of the environment of LDN 1780 on the Finkbeiner H$\alpha$ map reveals two giant H II regions at Galactic latitudes b $\sim +22\arcdeg$ just to the south of LDN 1780, with maximum H$\alpha$ surface brightnesses near $100\ R$. These two H II regions are powered by the OB stars $\zeta$ Oph and $\delta$ Sco, with approximate Hipparcos-derived distances \citep{Perryman+97} of 140 pc and 123 pc, respectively. These two H II regions are therefore separated from LDN 1780 by only about 50 pc to 40 pc to the south, illuminating LDN 1780 under albeit rather inefficient scattering angles near 70\arcdeg. The presence of these H II regions could well explain the greater relative strength of H$\alpha$ in the southern portions of LDN 1780. Considering the closeness of the fits in Fig. 3b, we conclude, in agreement with \citet{MJL07}, that both the absolute H$\alpha$ surface brightness in LDN 1780 and in particular its dependence on the optical depth through this cloud are fully consistent with dust scattering by typical interstellar grains in this cloud. These fits do not require a ten-fold locally-enhanced cosmic ray flux, nor are they consistent with the expected surface brightness distribution produced through photoionization by external Lyman continuum photons.

\section{CONSEQUENCES FOR THE HIGH-LATITUDE H$\alpha$ BACKGROUND}

The high-latitude Galactic sky is dominated by extended dust structures, which have been observed at both optical \citep{Sandage76, Witt+08} and far-infrared wavelengths \citep{Low+84}. The dust morphologies in the two spectral ranges mirror each other closely \citep{dVLP85, Stark95}, supporting the conclusion that the optical continuum intensity is due to scattering of the Galactic ISRF by wavelength-sized dust grains, while the far-infrared emission is due to the equilibrium thermal emission from the same dust grains, re-radiating that portion of the ISRF that is absorbed by them \citep{GT89, GC94}. Both the absolute and relative intensities of these two radiation phenomena are consistent with the intensity of the Galactic ISRF in the solar vicinity and the known scattering parameters \citep{Draine03} of interstellar dust.

At high latitudes, the intensities of the optical continuum and far-infrared backgrounds are linearly related, because the respective intensities result from scattering and absorption of the same interstellar radiation field by the same population of interstellar grains, and because the high-latitude sky is optically thin at both optical and far-infrared wavelengths. However, only at blue wavelengths can the optical background be fully attributed to scattering \citep{Mathis73}, while the optical background at red wavelengths appears to exhibit a significant contribution from dust luminescence or extended red emission (ERE), as shown by \citet{GT89} and \citet{GWF98}. Therefore, in the subsequent analysis we rely only on the observed relation between the intensities at blue optical and at far-infrared wavelengths.

For optically thin cirrus clouds, \citet{Witt+08} found a linear relation between the optical scattered light intensities near 457 nm and the 100 $\mu$m thermal emission intensities recorded by the IRAS survey \citep{SFD98}:

\begin{equation}
I (457\ nm)\ [\mathrm{erg\ cm^{-2} s^{-1} \AA^{-1} sr^{-1}}]  = 3.243 \times 10^{-9}\ I(100\ \mu m)\ [\mathrm{MJy\ sr^{-1}}]
\end{equation}

The optically thin high-latitude dust clouds exhibit a scattered light spectrum with an intensity ratio \citep{Witt+08}:

\begin{equation}
\frac{I (656\ nm)}{I(457\ nm)} \sim 0.68
\end{equation}

From the results shown in Fig. 3a,b and the associated curve fits, we adopt a line-to-continuum ratio
  
\begin{equation}
J(H\alpha) / J(656\ nm) = 6.0 \times 10^7\  [\mathrm{R\ erg^{-1}\ cm^{2}\ s^{1}\ \AA^{1} \ sr^{1}}].
\end{equation}

This corresponds to a fit to the H$\alpha$ intensities in Fig. 3b with a constant $A = 1.92$, which is intermediate to the two fits shown with $A = 1.7$ and $A = 2.35$ but closer to the lower one.

Combining the last three relations, we expect a scattered H$\alpha$ intensity in the high-latitude, optically thin dust corresponding to the measured far-infrared intensities $I $(100 $\mu m)\ [\mathrm{MJy\ sr^{-1}}]$:

\begin{equation}
I (H\alpha)[R]  = (0.129 \pm 0.015) \times I (100\ \mu m)\ [\mathrm{MJy\ sr^{-1}}].
\end{equation}

This relationship assumes that all high-latitude dust structures see an ISRF intensity approximately the same as that observed in the solar vicinity, with the same H$\alpha$ line to continuum ratio. We justify this assumption with the fact that virtually all high-latitude cirrus dust clouds owe their large angular extent and their high latitudes to their proximity to the sun \citep{Weiland+86, Franco89, SG99}. The cloud distances, typically 100 pc to 200 pc, are small in comparison with the distances of most of the principal H$\alpha$ sources at low Galactic latitudes.

By applying Eq. (6) to the $100\ \mu$m IRAS map of \citet{SFD98}, we now have a way for estimating the contribution to the high-latitude H$\alpha$ background due to Galactic dust scattering. We estimate the uncertainty in this prediction to be about 12$\%$, a combination of the estimated uncertainties in the relations shown in Eqs. (3), (4), and (5) of 3$\%$, 5$\%$, and 10$\%$, respectively.

We applied Eq. (6) to the entire high-latitude sky, where the linear, single-scattering approximation Eq.(6) is expected to be valid ($\tau(H\alpha) < 1$). Fig. 4a shows the resulting histogram of the numbers of $6\arcmin$ pixels with measured total H$\alpha$ intensities (black), predicted scattered H$\alpha$ intensities (green) and residual H$\alpha$ intensities (red) attributed to in-situ recombination in the WIM. The most likely value of scattered H$\alpha$ intensities (0.1 R) is about $19\%$ of the most frequently encountered measured total H$\alpha$ intensity (0.52 R). Thus, when averaged over the entire high-latitude sky, our prediction is in fair agreement with the expectations from radiative transfer models of \citet{WR99} and the recent analysis of the free-free emission and associated H$\alpha$ emission from the WIM by \citet{DD10}.

We next analyzed a northern Galactic sky region associated with extended cirrus structures, limited by Galactic longitudes ($340\arcdeg$ to $60\arcdeg$) and Galactic latitudes ($+35\arcdeg$ to $+75\arcdeg$). Fig. 4b shows the corresponding histograms of the frequency of the observed H$\alpha$ intensities from the map of \citet{Finkbeiner03}, of the predicted H$\alpha$ scattered light intensities, and the residual H$\alpha$ intensities that can be attributed to in-situ recombination, all without extinction corrections for this region. Here, the total observed H$\alpha$ intensity displays a distinct double-peaked distribution, while the predicted scattered H$\alpha$ intensities peak at a higher value (0.16 R) than for the high-latitude sky overall (0.10 R). We find that the removal of the scattered H$\alpha$ contribution leads to a narrowing of the frequency distribution of the residual H$\alpha$ intensities attributable to in-situ emission, implying smaller amplitude in the spatial structure of the in-situ emission component than previously thought \citep{Hill+07}, and an overall reduction in average intensity by about a factor of two. This implies that the free-free emission template for such high-latitude regions used in the WMAP reductions \citep{Bennett+03, Gold+10} has been overestimated in both strength and spatial structure. We suggest that the standard assumption in the reduction of CMB observations, namely that the observed H$\alpha$ all-sky map of \citet{Finkbeiner03} is a suitable template for free-free emission in the Galactic foreground, must be reexamined, especially as it applies to high-latitude parts of the sky.

In contrast to the effect of scattering, the correction for foreground extinction, using the standard approach of \citet{Bennett+03}, of the observed H$\alpha$ intensities or the H$\alpha$ residual intensities after removal of the scattered component introduces relatively small changes, as illustrated in Fig. 4c.

The observed H$\alpha$ high-latitude background before correction for scattered H$\alpha$ is much more structured than the component we identify as attributable to in-situ recombination. The FWHM values of the two respective intensity frequency distributions shown in Fig. 4c are different by about a factor of two. Fig. 5 shows a map of all high-latitude parts of the sky with 1 degree resolution, where the linear, single-scattering approximation Eq.(6) is expected to apply ($\tau(H\alpha) < 1$). We display the ratio of scattered H$\alpha$ to total observed H$\alpha$ intensities. We find many extended regions where the fraction of the scattered H$\alpha$ intensity is of order 0.5 or higher, while it is much lower in other regions. Specifically, at a resolution of 1 degree, we find for the unmasked areas, representing $81.4\%$ of the entire sky, that $51.7\%$ exhibits a ratio of scattered-to-total H$\alpha$ of $\le 0.25$, 38.7\% of the region falls into the range 0.25  - 0.50, $8.3\%$ reaches values of 0.5  - 0.75, with only $1.3\%$ exceeding the value of 0.75. The last category increases in coverage about fourfold when the pixel size is changed from 1 degree to 6\arcmin, with a corresponding decrease in the second interval (0.25 to 0.50). This emphasizes the fact that the scattered H$\alpha$ contribution adds particularly to the small-scale structure of the H$\alpha$ background.

The estimate by \citet{WR99}, suggesting that between 5\% and 20\% of the all-sky H$\alpha$ background may be attributable to scattering by dust appears to be in good agreement with our estimates for about half of the high-latitude sky, especially if lower latitudes are included (Fig. 4a), but the authors failed to account for the highly structured distribution in the ratio of H$\alpha$ scattering to H$\alpha$ in-situ emission (Fig. 4b) in their model. It is precisely this structure that is important in attempts to determine the small-scale anisotropy of the cosmic microwave background at angular scales of $\le$ 1 degree.

Also shown in Fig.~5 are the published positions where WHAM measurements of the [S II]/H$\alpha$ line intensity ratio have been obtained by \citet{Reynolds85, Reynolds88} and \citet{MRH06}. It appears, the selection criteria applied to these observations have resulted in a somewhat biased data set in the sense that the chosen directions fall either very close to the Galactic plane or predominantly in high-latitude regions where we identify the scattered H$\alpha$ contribution to be very small. We averaged the ratios of the scattered to total H$\alpha$ intensities predicted by Eq. (6) for all those individual pointings shown as small open circles falling outside the masked area in Fig. 5, with a resulting average of of scattered-to-total H$\alpha$ of 0.14. 
Also shown with a cyan outline is a continuous region surveyed by \citet{MRH06}. In Fig. 6 we show the line emission ratios [S II]/H$\alpha$ as a function of the predicted ratio of scattered-to-total H$\alpha$ for those pointings within this region that fall outside the near-plane masked area. For comparison, we also entered the corresponding values for the excess emission in LDN 1780, based on our recent WHAM measurements. While the data from the \citet{MRH06} survey appear to represent to a large extent a typical WIM environment with little scattering and significantly enhanced [S II]/H$\alpha$ ratios compared to those seen in H II regions, this comparison shows LDN 1780 to be clearly distinct from the WIM, with an emission line ratio similar to that of H II regions.This is an independent indication that the H$\alpha$ excess in LDN 1780 is a result of scattering. Note that the total ratio of scattered-to-total H$\alpha$ emission toward LDN 1780 is about 0.65, reflecting the fact that in addition to the H$\alpha$ excess in LDN 1780 analyzed in this paper, additional H$\alpha$ intensity of both scattering and recombination origin is seen in the direction of LDN1780.

It is our conclusion that the ratio of scattered to total H$\alpha$ background across the high-latitude sky is highly variable and that this could be demonstrated by conducting an unbiased survey of the [S~II]/H$\alpha$ line intensity ratio. While our results may lead to a reduction in the H$\alpha$ intensities attributable to in-situ emission in the WIM, they do not eliminate in-situ emission in the WIM as a real phenomenon requiring further study \citep[e.g.][]{Wood+10}.

After completion of this work a  paper by Seon et al. (arXiv:1006.4419) became available, dealing with the analysis of the far-ultraviolet (FUV) continuum background observed by SPEAR/FIMS. The latitude-dependent FUV background is consistent with scattering by interstellar dust. At Galactic latitudes $|b| > 25 \arcdeg$ these authors find a close linear correlation between the diffuse H$\alpha$ background taken from the Finkbeiner map and their FUV background intensities. The most likely interpretation of this result is that the high-latitude H$\alpha$ background is significantly affected by its dust-scattered component.


\section{CONCLUSIONS}
\begin{itemize}
\item We confirm the prediction of \citet{Jura79} and the findings of \citet{MJL07} that the excess surface brightness in H$\alpha$ light of LDN 1780, and by extension in other high-latitude dust structures \citep{LJM10}, is due to scattering of H$\alpha$ photons present in the Galactic ISRF, having originated in emission regions elsewhere.
\item Our determination of the line-to-continuum ratio of the scattered light in LDN 1780, when combined with earlier work relating the optical continuum intensity to the 100~$\mu$m surface brightness of optically thin high-latitude clouds, allows us to predict the intensity of the expected scattered H$\alpha$ intensity for large fractions of the high-latitude Galactic sky.
\item We find the ratio of scattered to total observed H$\alpha$ intensities for many regions dominated by Galactic cirrus dust to be of the order of 0.5, higher than predicted by \citet{WR99}, but with considerable small-scale structure and spatial variation.
\item The standard assumption in the reduction of CMB observations, namely that the observed H$\alpha$ all-sky map of \citet{Finkbeiner03} is a suitable template for free-free emission in the Galactic foreground, must be reexamined, especially as it applies to high-latitude parts of the sky.
\end{itemize}

\acknowledgements
We acknowledge constructive criticism and suggestions from two anonymous referees and fruitful discussions and sharing of information with Matt Haffner, Kalevi Mattila, C. R. O'Dell, and Kwang-Il Seon on subjects of emission line ratios and high-latitude scattering.
This research has been supported by grants from NASA and the NSF to the University of Toledo as well as by contributions from corporate sponsors AstroDon, RC Optical Systems, Santa Barbara Instrument Group, Software Bisque, and New Mexico Skies, for which we are grateful. CTD especially acknowledges the NSF-REU program at the University of Toledo, which supported the analysis of these data.


\clearpage

\begin{figure}
\centering{
\includegraphics[angle=270,width=3in]{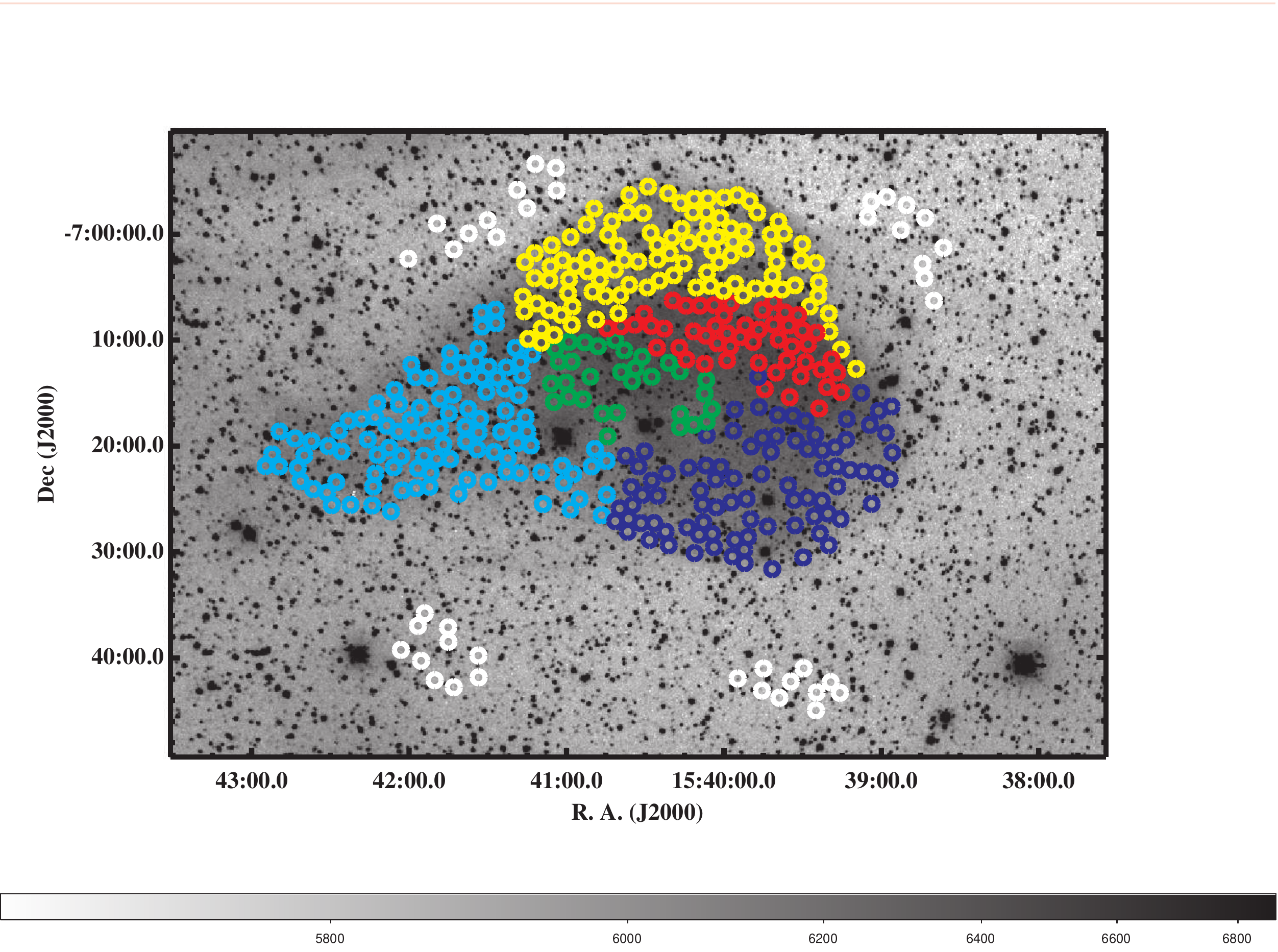}\\
\includegraphics[angle=270,width=3in]{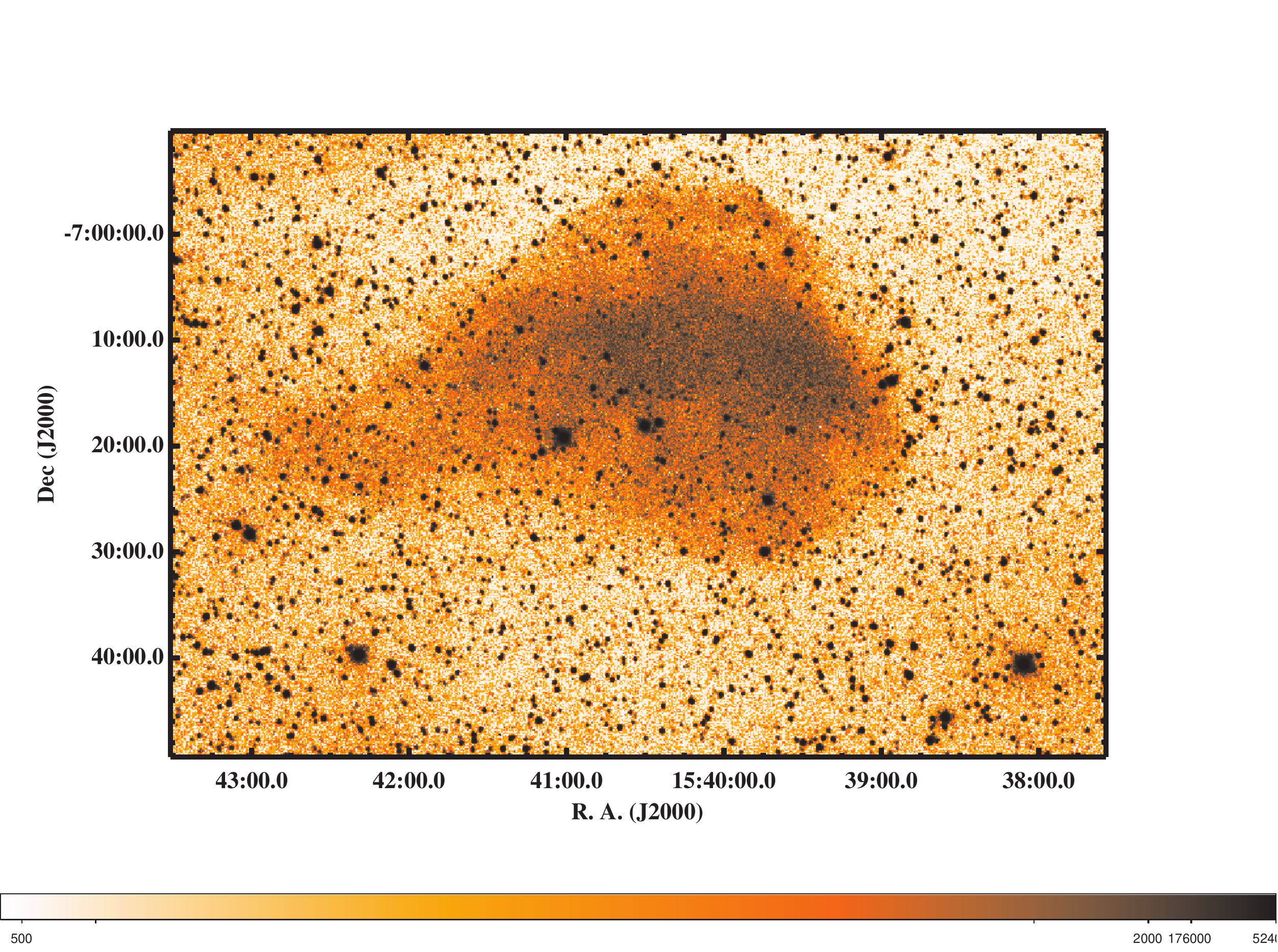}\\
\includegraphics[angle=270,width=3in]{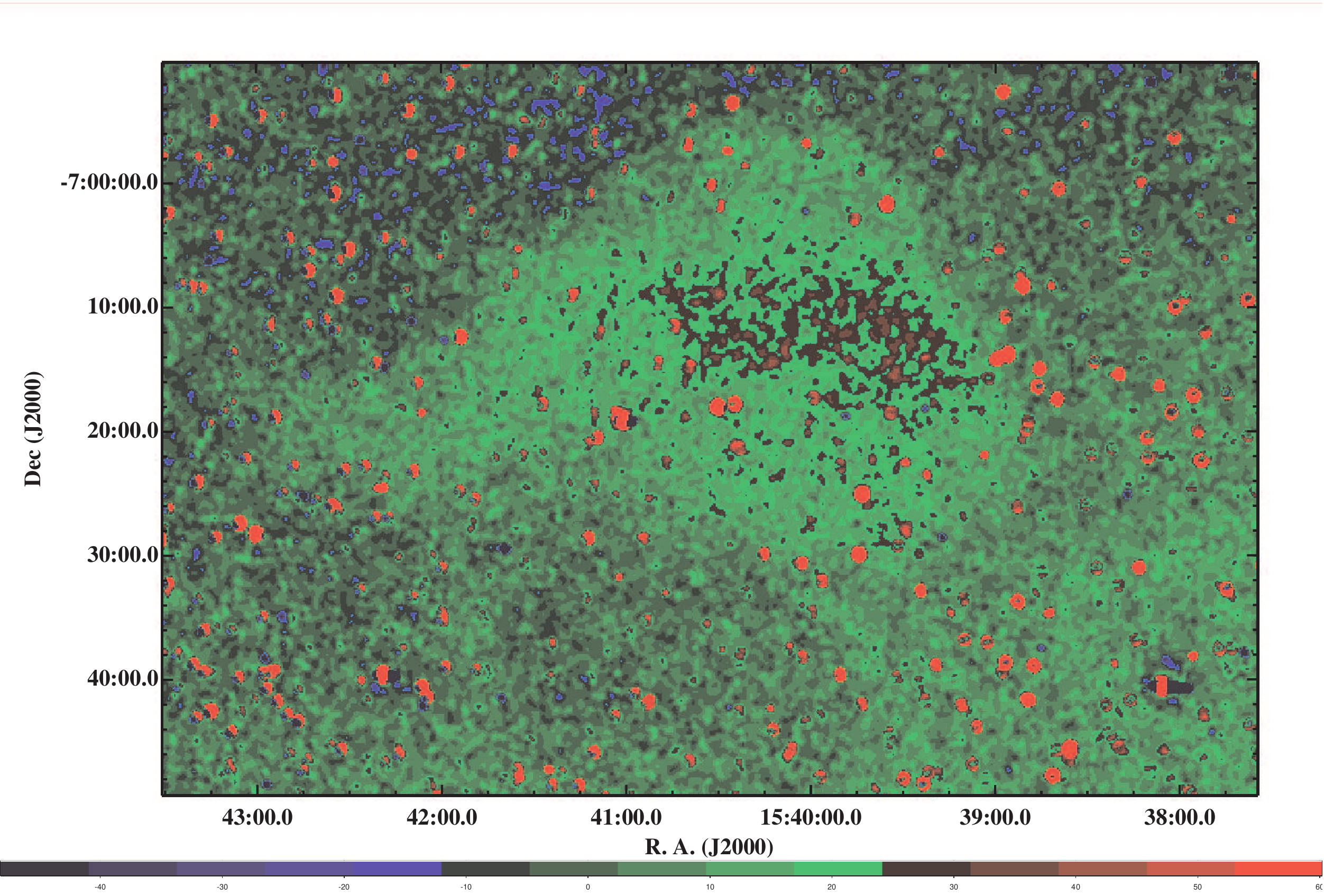}
\caption{(top)  Red continuum image of LDN 1780, identifying the 378 sampling areas (colored circles) and 40 sky reference fields (white). The field shown is $88\arcmin \times 56\arcmin$. Each sampling area is $74\arcsec$ in diameter. Typical line-of-sight extinctions through LDN 1780 are  A$_V \sim 1$ mag (cyan, blue, yellow), A$_V \sim 2$ mag (green) and A$_V \sim 4$ mag (red). North is up and east is to the left.  (middle) 9000 s exposure of LDN 1780 with a 6 nm wide narrow-band H$\alpha$ filter.  (bottom)  H$\alpha$ excess image of LDN 1780 after corrections for sky and continuum contributions in the narrow-band H$\alpha$ filter. The image has been smoothed with a 7-pixel Gaussian. The intensity scales are counts px$^{-1}$.}
}
\end{figure}

\begin{figure}
\centering{
\includegraphics[width=6in]{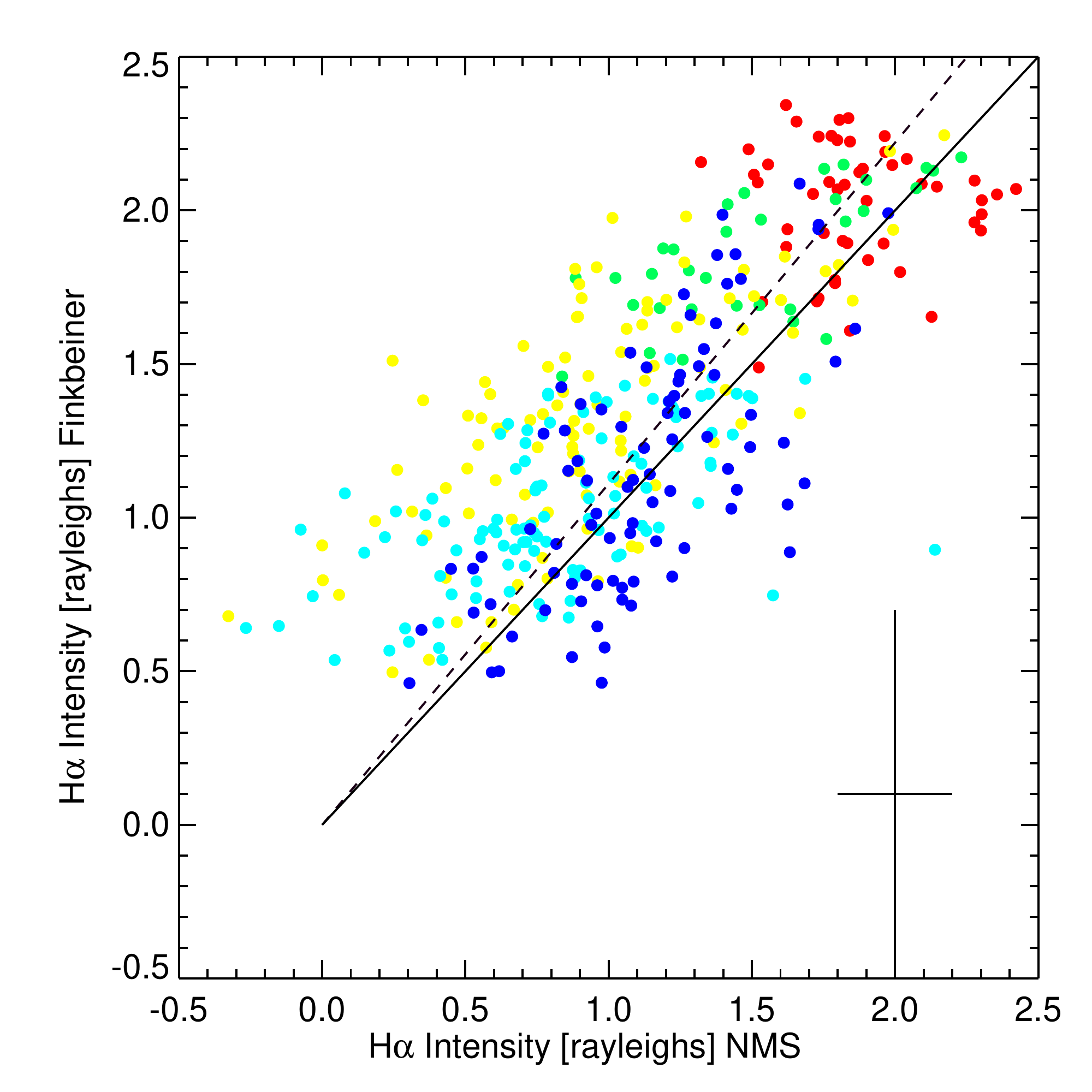}
\caption{Cross correlation of H$\alpha$ surface brightnesses in LDN 1780, measured on our New Mexico Skies (NMS) image vs. the \citet{Finkbeiner03} map. The symbol colors refer to the source areas identified in Fig. 1(top). The solid diagonal line indicates the expected 1:1 correlation; the dashed curve is a linear least-squares fit to the data. On average, the \citet{Finkbeiner03} intensities are about 11\% higher.}
}
\end{figure}

\begin{figure}
\centering{
\includegraphics[width=3in]{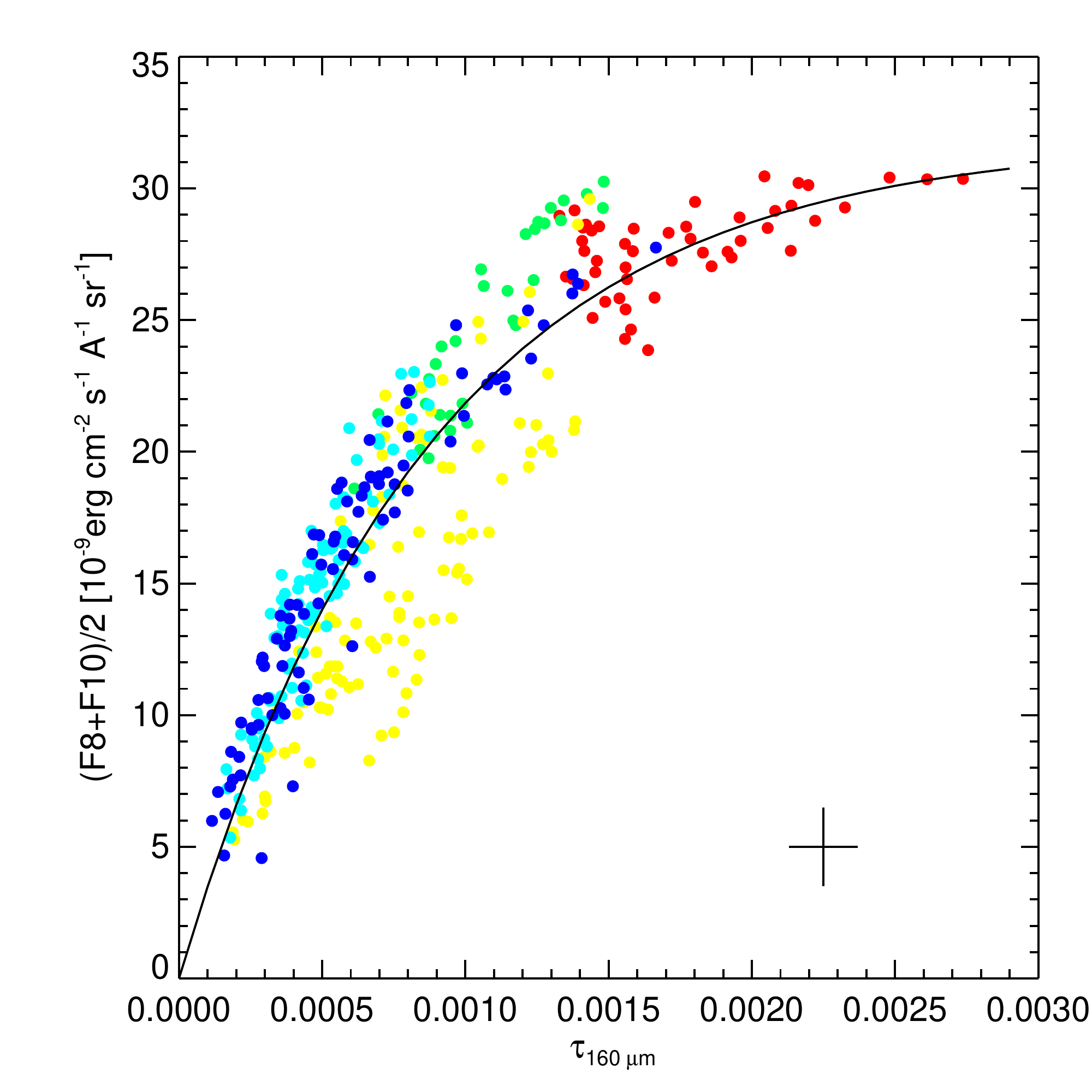} \includegraphics[width=3in]{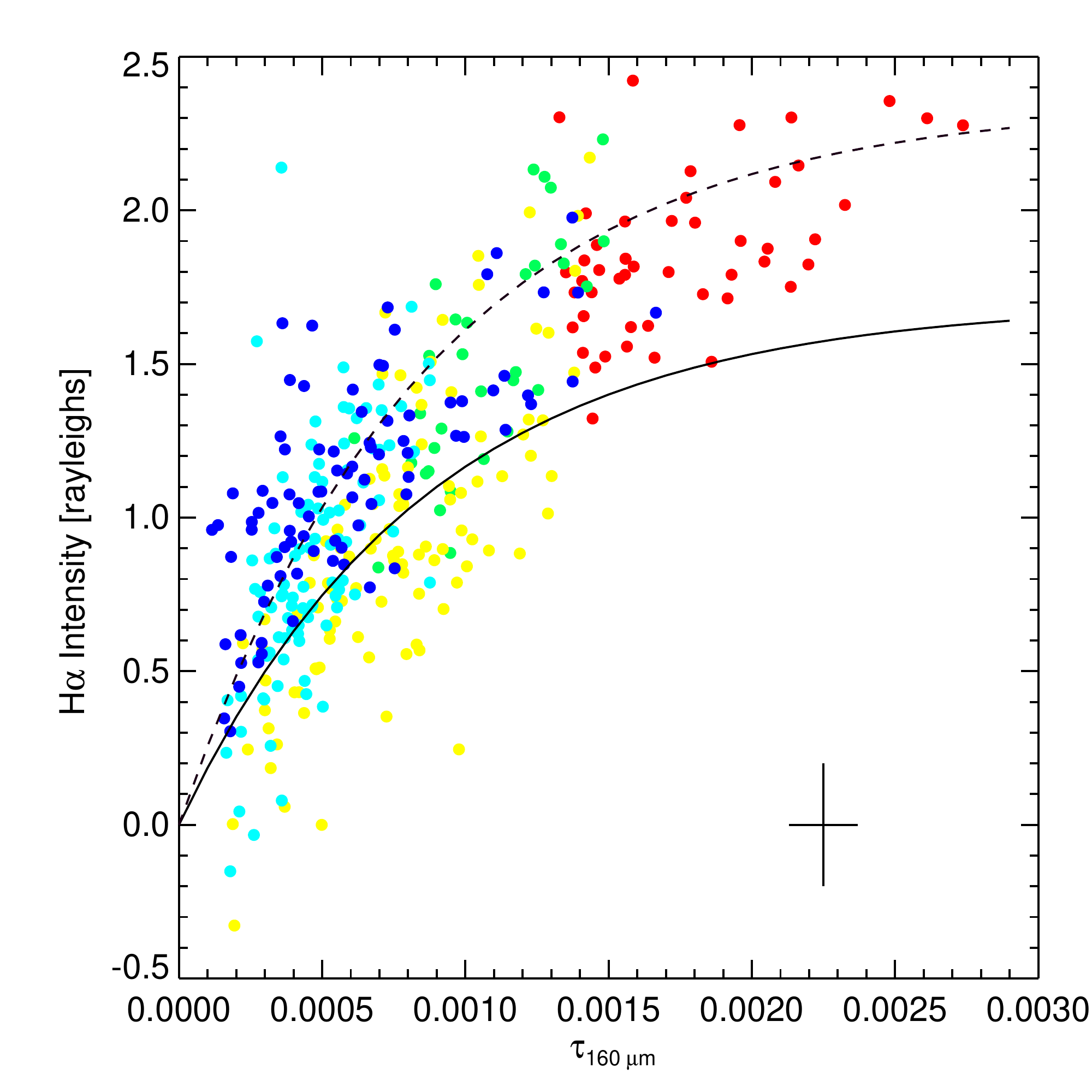}
\caption{(a) Averaged continuum intensities in LDN 1780, measured with the BATC \#8 and BATC \#10 filters, are plotted as a function of the optical depth at 160 $\mu$m ($A_V = 1$ mag corresponds to $\tau (160)\ \mu$m) = 0.0006). The symbol colors refer to the different source regions identified in Fig. 1(top). The solid curve represents the least-squares fit of the function given in Eq. (2) to the data. (b) Plot of the excess H$\alpha$ intensity (NMS data) in LDN 1780 as a function of the optical depth at 160 $\mu$m . The symbol colors refer to the different source regions defined in Fig. 1 (top). The solid curve shows the expected scattered light intensities for the case here the H$\alpha$ line-to-continuum ratio is the same as observed at the sun. The dashed curve represents the case where the H$\alpha$ line intensity is increased by 38\% compared to the continuum relative to the solar neighborhood.}
} 
\end{figure}

\begin{figure}
\centering{
\includegraphics[width=3in]{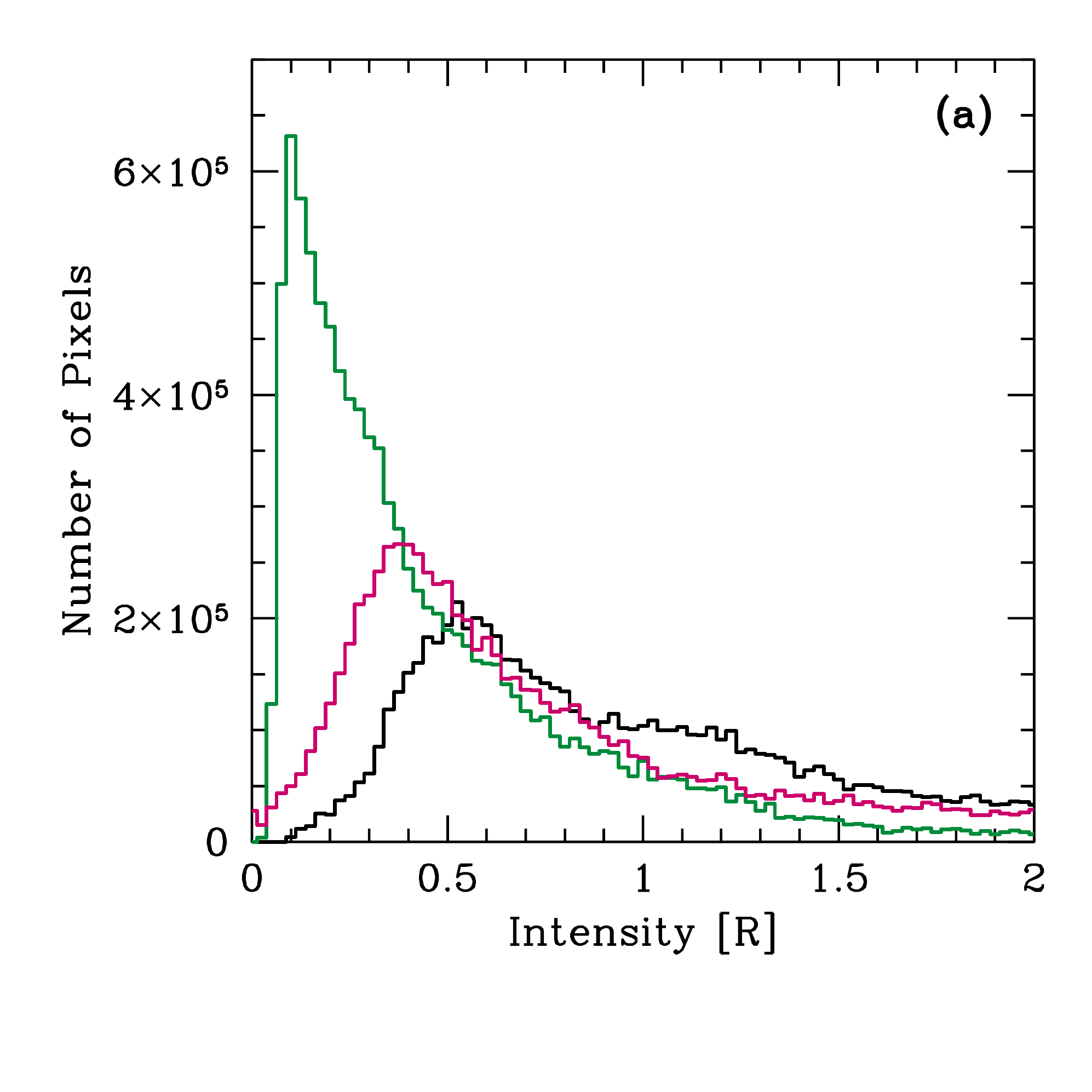}
\includegraphics[width=3in]{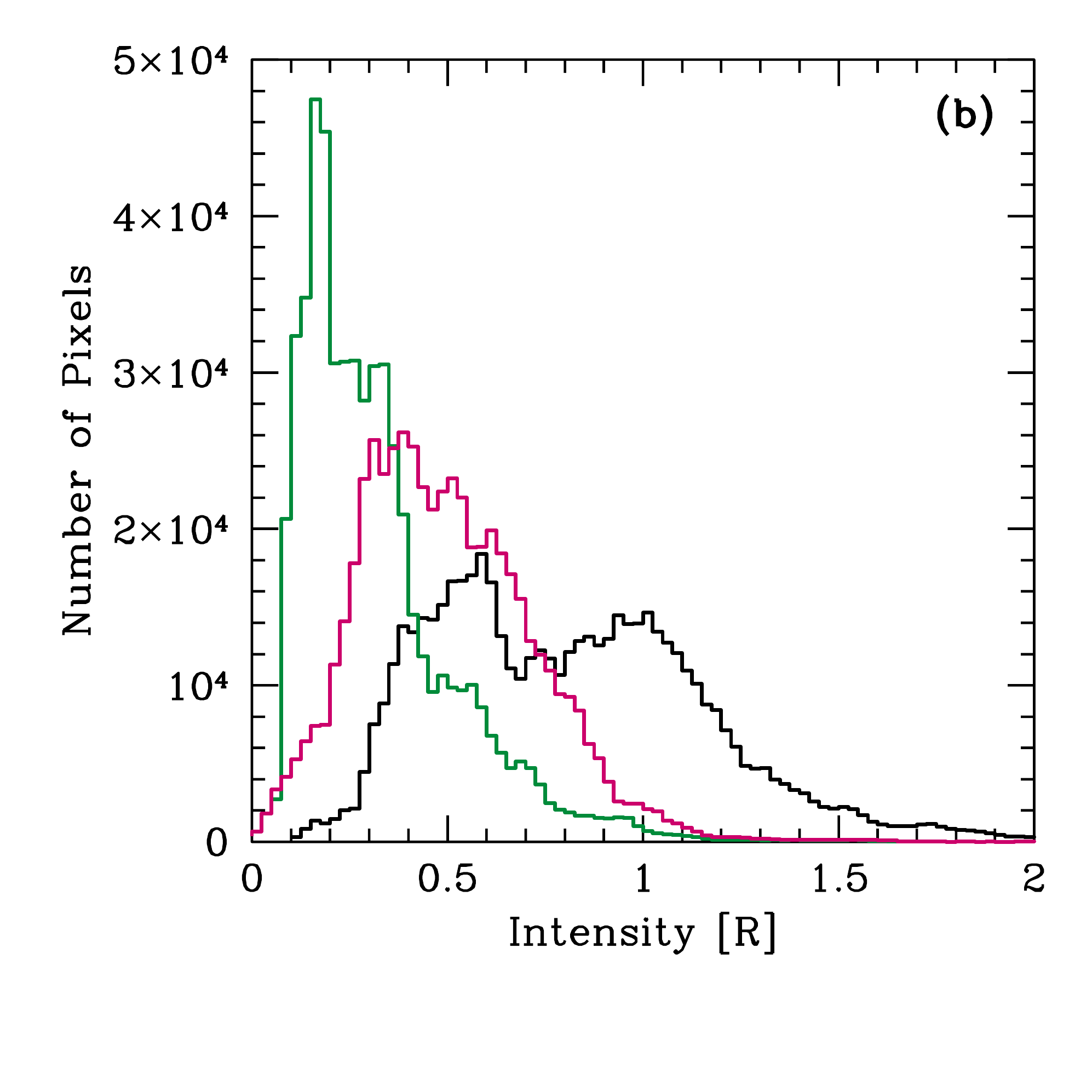}
\includegraphics[width=3in]{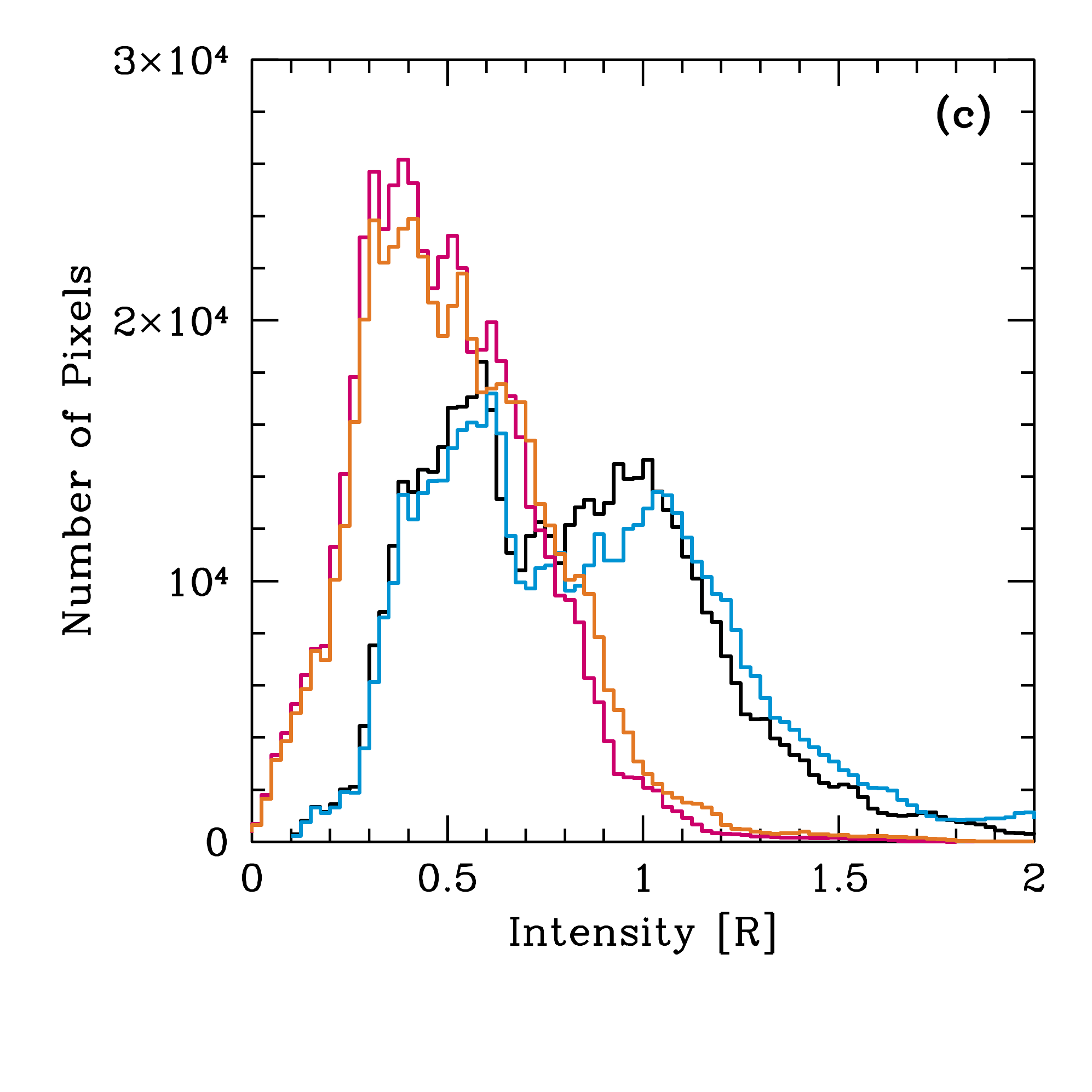}
\caption{(a) Frequency distributions of total observed H$\alpha$ intensities (black) from the \citet{Finkbeiner03} map for the entire high-latitude sky limited by the condition $\tau(H\alpha) < 1$, of the predicted scattered H$\alpha$ component (green) according to Eq. (6), and of the difference (red), which we attribute to in-situ recombination emission.---(b) Frequency distributions of total observed H$\alpha$ intensities (black) from the \citet{Finkbeiner03} map for a high-latitude region bounded by l = $340 \arcdeg$ and l = $60 \arcdeg$ on longitude, b = $+35 \arcdeg$ and b = $+75 \arcdeg$ in latitude, of the predicted scattered H$\alpha$ component (green) according to Eq. (6), and of the difference (red), which we attribute to in-situ recombination emission. ---(c) Frequency distributions of the total observed H$\alpha$ intensities before (black) and after (blue) extinction correction according to \citet{Bennett+03} compared with the residual H$\alpha$ intensities after subtraction of the scattered light component, again before (red) and after (pink) extinction correction.}
}
\end{figure}

\begin{figure}
\centering{
\includegraphics[width=5in]{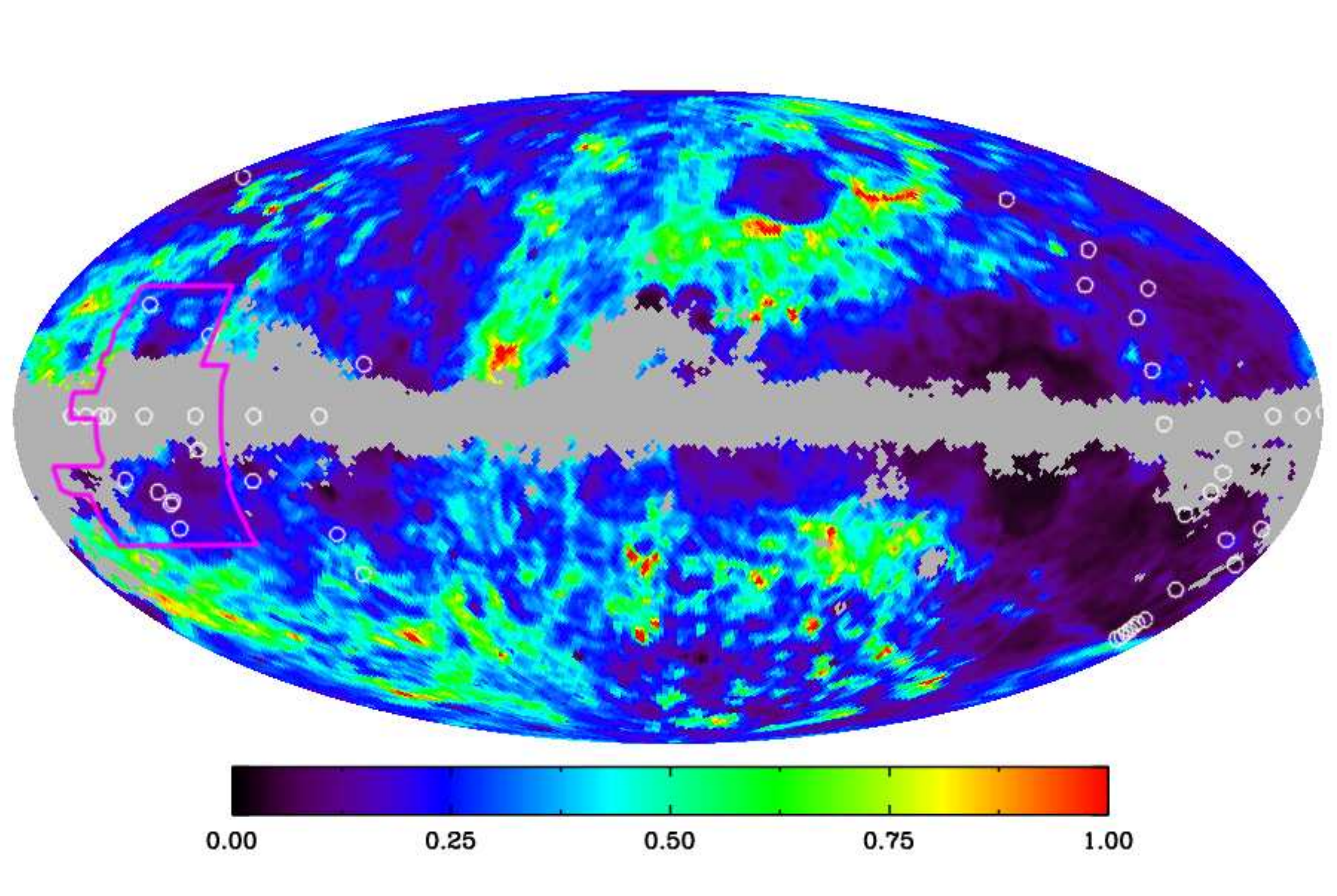}
\caption{An all-sky map (1$\arcdeg$ resolution map) of the ratio of the predicted scattered H$\alpha$ intensities to the total observed H$\alpha$ intensities from the \citet{Finkbeiner03} map. The blocked-out regions of the map are areas where the optically thin condition $\tau(H\alpha) < 1$ is no longer valid. The small white circles identify the published positions of WHAM measurements of the [S~II]/H$\alpha$ emission line ratio by \citet{Reynolds85, Reynolds88} and \citet{MRH06}. The region outlined in cyan represents the portion of the sky in which detailed surveys of emission line ratios were conducted by \citet{MRH06}.}
}
\end{figure}

\begin{figure}
\centering{
\includegraphics[width=5in]{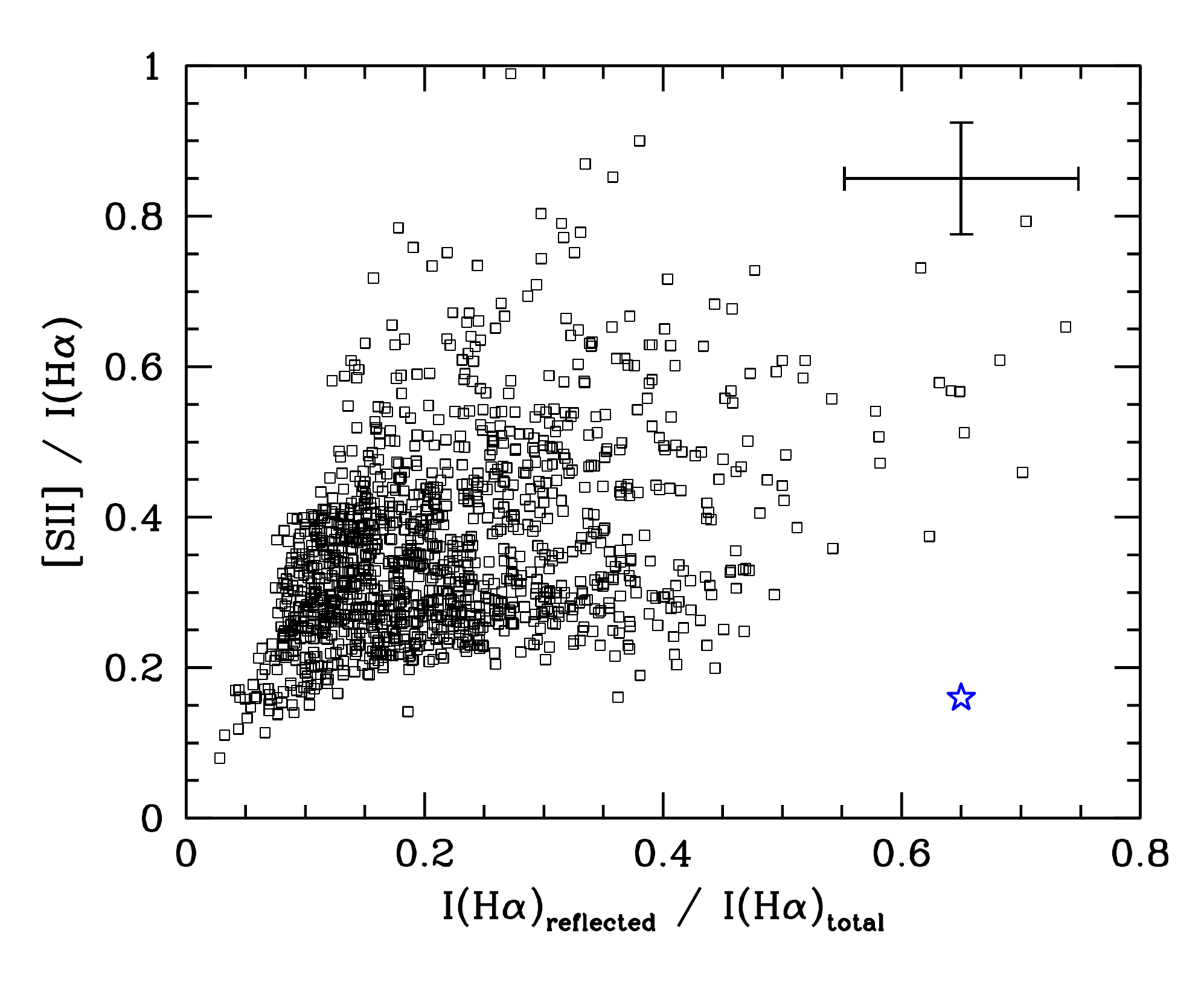}
\caption{Scatter diagram of individual measurements of the line emission ratio [S II] / H$\alpha$ in the \citet{MRH06} survey field identified in Fig. 5 plotted against the predicted ratios of scattered-to-total H$\alpha$ intensity from our 1$\arcdeg$-resolution map shown in Fig. 5. The correspondingly measured values for LDN 1780 are indicated by a star symbol. Note that the line emission ratio [S II] / H$\alpha$ for LDN 1780 is the result of differential spectrophotometry characterizing the excess emission in LDN 1780 relative to the surrounding sky regions, while the \citet{MRH06} survey field data are the result of absolute photometry.}
}
\end{figure}

\begin{deluxetable}{llccr}
\tablecaption{Filter Characteristics and Exposures}
\tablehead{\colhead{Filter} &\colhead{$\lambda$ [nm]} & \colhead{FWHM [nm]} & \colhead{Exp. Time [s]} & \colhead{Obs. Date}}
\startdata
BATC\#8	&	607	&	30.0	&	9000	&	4 April, 2008\\
BATC\#9	&	666	&	51.7	&	8100	&	4 April, 2008\\
BATC\#10	&	705	&	29.0	&	9000	&	4 April, 2008\\
H$\alpha$ Narrow	&	656.3	&	6.0	&	9000	&	11 April, 2008\\
\enddata
\end{deluxetable}

\end{document}